\documentclass[aps,pra,superscriptaddress,floatfix,reprint,showkeys,showpacs]{revtex4-1}

\usepackage{graphicx}
\usepackage{subfigure}
\usepackage{float}
\usepackage[space]{grffile}
\usepackage{leftidx}
\usepackage{amsmath,amssymb,amstext}
\usepackage{xcolor}
\usepackage{wrapfig}
\usepackage{hhline}
\usepackage{xcolor}
\usepackage{hyperref}
\usepackage{amsthm}
\usepackage{footnote}

\def\Rb{$^{87}$Rb}
\def\Li{$^{7}$Li}

\def\m1cond{$m_F{=}-1$}
\begin{document}
\title{Cross-dimensional relaxation of $^7$Li-$^{87}$Rb atomic gas mixtures in a spherical-quadrupole magnetic trap }
\author{Fang Fang}
\affiliation{Department of Physics, University of California, Berkeley, California 94720, USA}
\email{akiraff@berkeley.edu}
\author{Shun Wu}
\affiliation{Hubei Key Laboratory of Optical information and Pattern Recognition, Wuhan Institute of Technology, Wuhan 430205, China}
\email{wushun@wit.edu.cn}
\author{Aaron Smull}
\affiliation{Department of Physics, University of California, Berkeley, California 94720, USA}
\author{Joshua A. Isaacs}
\affiliation{Department of Physics, University of California, Berkeley, California 94720, USA}
\author{Yimeng Wang}
\affiliation{Department of Physics and Astronomy, Purdue University, West Lafayette, Indiana 47907, USA }
\author{Chris H. Greene}
\affiliation{Department of Physics and Astronomy, Purdue University, West Lafayette, Indiana 47907, USA }
\affiliation{Purdue Quantum Science and Engineering Institute, Purdue University, West Lafayette, Indiana 47907, USA }
\author{Dan M. Stamper-Kurn}
\affiliation{Department of Physics, University of California, Berkeley, California 94720, USA}
\affiliation{Materials Sciences Division, Lawrence Berkeley National Laboratory, Berkeley, California 94720, USA}

\begin{abstract}
We measure the interspecies interaction strength between $^7$Li and $^{87}$Rb atoms through cross-dimensional relaxation of two-element gas mixtures trapped in a spherical quadrupole magnetic trap.  We record the relaxation of an initial momentum-space anisotropy in a lithium gas when co-trapped with rubidium atoms, with both species in the $|F=1, m_F = -1\rangle$ hyperfine state.  Our measurements are calibrated by observing cross-dimensional relaxation of a \Rb-only trapped gas.  Through Monte Carlo simulations, we compare the observed relaxation to that expected given the theoretically predicted energy-dependent differential cross section for \Li-\Rb\ collisions.  The experimentally observed relaxation occurs significantly faster than predicted theoretically, a deviation that appears incompatible with other experimental data characterising the \Li-\Rb\ molecular potential.

\end{abstract}
\maketitle

In ultracold atomic gas experiments, a detailed and specific characterization of collisions within the gas is crucial to controlling the gas and understanding its dynamics and thermodynamics.  This characterization is achieved by collating diverse sources of information on the interatomic potential, e.g.\ from direct measurement of the interatomic collision cross section and of the interatomic interaction energy in ultracold regime, from photoassociation spectroscopy, and from the observation of magnetic-field-induced Fano-Feshbach resonances \cite{wein99rmp,wein03collisionbook}.  Given the low temperature at which atom-atom collisions occur, a characterization of just the lowest-energy partial waves, is sufficient to guide and interpret experiments.  An increasing number of experiments study quantum gas mixtures, in which atoms of more than one element or isotope are held in the same trap and collide with one another.  Each combination requires a new set of experiments to characterize the particular collisions within each gas.

In this work, we consider cold collisions between $^7$Li and $^{87}$Rb atoms, with collision energies in the range of several 100 $\mu$K.  Low energy collisions between these elements have previously been characterized by observing thermalization between Li and Rb gases with both atoms held in the magnetically trapped $|F=2, m_F = 2\rangle$ hyperfine ground state \cite{marz07therm}, via which the magnitude of the triplet scattering length for $^7$Li-$^{87}$Rb collisions could be inferred.  Previous measurements of collisional Fano-Feshbach resonances between atoms both in the $|F=1, m_F = +1\rangle$ hyperfine state, observed at large magnetic fields \cite{marz09,maie15efimov}, together with photoassociation \cite{dutt13epl,dutt14lirb,stev16lirb} and Raman spectroscopy \cite{ivan11lirb} of several isotopologues of the LiRb molecule have led to the development of an empirically based model of the LiRb molecular potentials \cite{maie15efimov}, from which the low-energy scattering amplitudes can in principle be predicted for all hyperfine-state combinations undergoing low-temperature collisions.

In our experiment, we prepare laser cooled gases of both $^7$Li and $^{87}$Rb and load them with an anisotropic momentum-space distribution within a spherical-quadrupole magnetic trap with both species of atoms being polarized in the $|F=1, m_F = -1\rangle$ state.  We then observe the cross-dimensional relaxation\cite{CDRClarify} of the momentum-space distribution after a variable interaction time.  The experimentally measured timescale for cross-dimensional relaxation is compared to a numerical model in which collisions are assumed either to occur through an energy-independent, isotropic cross-section, or through the energy-dependent differential cross-section predicted theoretically \cite{maie15efimov}.  Our experimental results are calibrated by performing experiments both with Li-Rb mixtures and also with a Rb-only gas.  We observe the Li-Rb cross-dimensional relaxation to occur faster than predicted theoretically, implying that the theoretical models underestimate the collision cross section in the energy range probed by our experiment.

Our measurements are the first to test these model predictions for collisions between $^7$Li and $^{87}$Rb atoms, with each atom in its magnetically trapped $|F=1, m_F = -1\rangle$ hyperfine state (as opposed to the $|F=2, m_F = 2\rangle$ state \cite{marz07therm}) at near-zero magnetic field.  Such measurements are vital for assessing the prospects for sympathetic evaporative cooling of gases in these trapped states.  Testing the model predictions also leads to better predictions for the nature of heteronuclear spinor Bose-Einstein gases composed of both $^7$Li and $^{87}$Rb gases in the $F=1$ spin manifolds.  Novel phenomena in heteronuclear spinor Bose-Einstein gases of $^{23}$Na and $^{87}$Rb gases have been recently reported \cite{li15heter}, in which spin-dependent interactions between two spinor Bose-Einstein gases lead to coherent exchange of spin polarization between them.  The interaction strength measured in the present work, being sensitive to both the electron singlet and triplet molecular potentials, can help improve our knowledge of the spin-dependent interactions in the $F=1$ $^7$Li-$^{87}$Rb system.

Our paper is structured as follows. Sec.\ \ref{sec:preparation} describes our experimental setup for laser-cooling gases of lithium and rubidium, trapping them magnetically, and then measuring their momentum distributions.  In Sec.\ \ref{sec:measurements}, we describe our observations of cross-dimensional relaxation of atomic gases that evolve within the spherical-quadrupole magnetic trap.  We characterize the evolution of three gas samples: Observations on a lithium-only gas, exhibiting very weak interatomic interactions, confirm the fact that motion in a spherical quadrupole trap is non-ergodic in the absence of collisions.  Observations on a rubidium-only gas, whose stronger interatomic interactions are well characterized \cite{egor13}, show rapid cross-dimensional relaxation and are used to calibrate our experimental method.  Finally, observations of the cross-dimensional relaxation of the lithium momentum distribution within a lithium-rubidium gas mixture serve as a measurement of the Li-Rb interaction strength.

Sec.\ \ref{sec:modeling} discusses how measurements of the cross-dimensional relaxation rate are related to predictions for the energy-dependent cross-section for elastic collisions.   We derive analytic expressions that relate the measured relaxation rate to a ``cross-dimensional relaxation cross-section,'' $\sigma_\mathrm{cdr}$, if we assume such cross-section to be isotropic and energy independent.  These analytic expressions allow for rough comparisons between the experimentally measured relaxation rate and the theoretically predicted cross section, and for the propagation of experimental and modeling errors.  More precise experiment/theory comparisons are performed using Monte Carlo simulations that account for energy dependence and anisotropy in the differential cross section, and also for modifications of the magnetic trap potential that occur at high magnetic fields.  The conclusion of this precise comparison is that our experimental measurements indicate the collision cross-section between $^7$Li and $^{87}$Rb atoms to be larger than predicted, within the energy range probed in our experiment.

In Sec.\ \ref{sec:theory}, we attempt to reconcile the tension between our experimental findings and theoretical models of $^7$Li-$^{87}$Rb atomic interactions.  We explicate how theoretical models of the $^7$Li-$^{87}$Rb molecular potential are constrained by several prior measurements, most significantly by measured positions of several Fano-Feshbach resonances \cite{marz09,maie15efimov}.  Our conclusion is that the discrepancy between our experimental findings and the predictions of theoretical models remains significant and unresolved.

\section{Preparation of lithium-rubidium laser-cooled gas mixtures}
\label{sec:preparation}

For this work, laser-cooled gas mixtures of $^7$Li and $^{87}$Rb were produced using a Zeeman slower, a magneto-optical trap (MOT), and sub-Doppler laser cooling.  A two-element Zeeman-slowed atomic beam was produced by first generating a collimated effusive atomic beam from a dual-species effusive oven.  In this oven, a lower-temperature chamber, containing metallic rubidium and maintained at around 150$^\circ$C, emits a Rb atomic flux into a higher-temperature chamber, containing metallic lithium and maintained at around $500^\circ$C.  The two gases mix within this high-temperature oven, and then propagate through a capillary-array nozzle into a high vacuum chamber.  The atoms pass through an additional differential-pumping chamber before entering the Zeeman slower.  The Zeeman slower, described in Ref.\ \cite{mart10slower}, is designed to decelerate both Li and Rb atoms simultaneously.  However, in the present work, the slower was operated for each of the elements sequentially.

The Zeeman-slowed beam was trapped within a two-species magneto-optical trap.  In each experimental cycle, the trap was initially loaded just with Li atoms, which were produced by sending only the lithium-slowing light through the increasing-field Zeeman slower and tuning the slower and the magneto-optical trap magnetic fields so as to optimize the lithium MOT loading rate.  After loading the lithium MOT for a variable time, we switched the Zeeman slower field and illumination to produce a Zeeman-slowed Rb beam.  Similar to Refs.\ \cite{lado09,dutt14losses}, we observe Rb atoms and Rb-cooling light to cause strong light-induced losses of lithium atoms from the MOT.  To mitigate these losses, we displaced the center of the Rb MOT by applying an additional ``pusher'' light beam to produce an imbalanced radiation pressure force onto the rubidium atoms while the MOT operates for both elements.  Following the loading of both species into the MOT, we extinguished the pusher beam over 20 ms to allow the two elements to overlap within the MOT.  We then compressed both MOTs for 2 ms to increase the density of both species \ \cite{Lewandowski2003}. The MOT magnetic field was then rapidly turned off.

Following a brief delay to allow residual eddy fields to decay, we applied a 2-ms pulse of D1 gray molasses and dark-state cooling to the Li atoms \cite{grie13D1,siev15}.  For this, along all directions of the previously applied Li MOT beams, we insert light that is $2 \pi \times 35$ MHz blue-detuned from the $|F=2\rangle \to |F^\prime=2\rangle$ transition on the D1 line.  On this light, we add also an 802.7 MHz frequency sideband, using an electro-optical modulator, with a power of 4\% of the carrier beam.  This sideband drives a two-photon resonance between the \Li\ ground hyperfine states, creating a velocity-dependent dark state into which the atoms are cooled.  Through this method, we cooled almost all lithium atoms from the compressed MOT to a final temperature of around 40 $\mu$K in all directions.  During this time, rubidium atoms were subject to polarization gradient cooling, and brought thereby to a similar temperature as the lithium gas.

The laser-cooled gas was then trapped within a spherical-quadrupole magnetic trap, formed by sending a large current through the coils previously used for the MOT.  The number of trapped atoms was enhanced by applying optical pumping to both elements just before magnetic trapping.  We trap as many as $6 \times 10^8$ Rb atoms and $2 \times 10^7$ Li atoms within the magnetic trap.  For several of the measurements reported here, the atom number in each element was adjusted by reducing the MOT loading times appropriately, or by changing the composition of the effusive atomic beam by adjusting temperatures in the two-element oven.


Important to this work, the atomic momentum distribution in the spherical-quadrupole trap is initially anisotropic.  This anisotropy arises from a mismatch between the spherical or slightly prolate spatial distribution of the atoms at the end of laser cooling, and the oblate equipotential lines of the spherical-quadrupole trap.  As a result of this mismatch, following a few oscillation periods in the magnetic trap, the gases evolve to a quasi steady-state distribution in which the ensemble-averaged kinetic energy $E_z$ along the axial trap direction is larger than that along the radial directions ($E_x$ and $E_y$).  Any initial center-of-mass motion of the magnetically trapped gases is also dissipated within a few periods of motion.

The atom number and momentum distribution for each element is measured by switching off the magnetic trap, allowing the gas to expand, and then imaging its distribution with resonant absorption imaging.  To relate the observed distribution accurately to the momentum distribution of the trapped gas, it is necessary to account for the non-instantaneous switch-off of the magnetic trapping field; we observe that the magnetic field is effectively ramped off over around 1.5 ms, a timescale that is set by electronic components in our setup.  This gradual switch-off  has a particularly significant impact on the propagation of the lithium gas, which is allowed a total time of flight of only up to 5 ms.  To account for this impact, we performed numerical simulations of classical motion in the spherical-quadrupole trap as it is ramped down, allowing us to relate the position-space distribution after time of flight to the momentum-space distribution before the trap is turned off.

\section{Cross-dimensional relaxation measurements}
\label{sec:measurements}


Cross-dimensional relaxation has been used to determine the collision properties of single-element \cite{monr93cscs} and two-element atomic gases \cite{bloc01sympathetic,gold04}.  These experiments have typically been performed on atoms trapped in harmonic potentials, generated either magnetically or optically. In such potentials, the motion of non-interacting atoms separates into decoupled equations of motion in three directions corresponding to the principal axes of the trap. Therefore, a distribution of non-interacting atoms prepared initially with an energy anisotrosopy in each principal direction of motion will retain that anisotrosopy indefinitely. Collisions cause the energy to be redistributed, relaxing toward a thermal equilibrium distribution.


Our experiment differs slightly from previous work in that it is performed with atoms trapped in a spherical-quadrupole magnetic trap, with a total field $B(\textbf{r}) = B'(\frac{x}{2},\frac{y}{2},-z)$. Neglecting geometric effects of the motion of spins in an inhomogeneous magnetic field and non-adiabatic Majorana loss effects \cite{majo32,suku97}, atoms at low energy move in a potential of the form
\begin{equation}
U(x,y,z)_\mathrm{lin} = \mu B^\prime \sqrt{z^2 + (x^2 + y^2)/4}
\end{equation}
where $x$, $y$ and $z$ are Cartesian coordinates, $B^\prime$ is the magnetic field gradient along the trap axis, and $\mu = \mu_B/2$ is the atomic magnetic moment at low field in the $|F=1, m_F = -1\rangle$ hyperfine state, with $\mu_B$ being the Bohr magneton.  We neglect the small nuclear magnetic moments. This expression for the potential is accurate only in the regime where the linear Zeeman shift is far smaller than the atomic hyperfine splitting.  Outside this regime, a more accurate form of the potential can be derived using the Breit-Rabi formula \cite{brei31}.  This intermediate-field correction to the magnetic trap potential is relevant to our trapped lithium gas, given the small hyperfine splitting in the \Li\ ground state.  We account for the full Breit-Rabi expression for the magnetic potential in our Monte Carlo simulations.  In contrast, for our trapped rubidium gas, the linear Zeeman energy expression is sufficiently accurate\cite{CloudSizeClarify}.


While the equations of motion are not plainly separable, it is observed through numerical simulation that single particle motion in the spherical-quadrupole potential is quasi-periodic and non-ergodic \cite{gome97}.  As a result, in the absence of collisions, an anisotropy in the momentum-space distribution of an atomic gas trapped in such a spherical-quadrupole trap is expected to persist indefinitely. 

We confirm this expectation experimentally by observing a trapped gas containing just bosonic \Li.  We produce this gas using the method outlined above, while skipping the rubidium loading cycle and turning off all rubidium-cooling light.  The initial anisotropy in the ensemble- and cycle-averaged kinetic energies of the trapped atoms is maintained for as long as 40 seconds, approaching the vacuum-limited magnetic trap lifetime.   The interactions between \Li\ atoms in the $|F=1, m_F = -1\rangle$ state are extremely weak, characterized by an s-wave scattering length $a_\mathrm{Li,Li}$ of just 5 $a_B$ \cite{stre02}, and thus a collision cross section of just $8 \pi a_\mathrm{Li,Li}^2 = 1.8 \times 10^{-14} \, \mbox{cm}^{2}$.  Therefore, at the density ($n_\mathrm{Li} \sim 10^{9} \, \mbox{cm}^{-3}$ at the trap center) and kinetic temperatures ($T_{x,z} = 2 E_{x,z}/k_B \sim 290 \, \mu{K}$ in each direction) of the trapped Li gas, the atoms collide with a per-atom rate of just $\Gamma_\mathrm{Li,Li} =0.002 \, \mbox{s}^{-1}$.  Thus, the atoms are essentially non-interacting, and their motion non-ergodic, over the experimental time scale. Non-ergodic motion in a spherical quadrupole trap was observed previously in experiments using spin-polarized $^6$Li atoms, which, being fermions, are strictly non-interacting at the zero collisional energy limit\cite{Li6K40thesis, Suchet_2016}.


By comparison, in the presence of collisions, the initially anisotropic momentum distribution is observed to relax toward an isotropic distribution.  We observe this cross-dimensional relaxation in two variations of our experiment.  In the first variation, we observe cross-dimensional relaxation of a gas composed just of Rb atoms.  The strength of Rb-Rb collisions is already well determined \cite{bugg04dwave}; thus, measurements of cross-dimensional relaxation in this single-element gas allow us to confirm and calibrate our experimental method.  In the second variation, we observe cross-dimensional relaxation of the momentum-space distribution of Li atoms in the presence of Rb atoms.  The relaxation of the lithium distribution can be ascribed completely to Li-Rb collisions.

\begin{figure}[t]
	\begin{center}
		\includegraphics[width=0.45\textwidth]{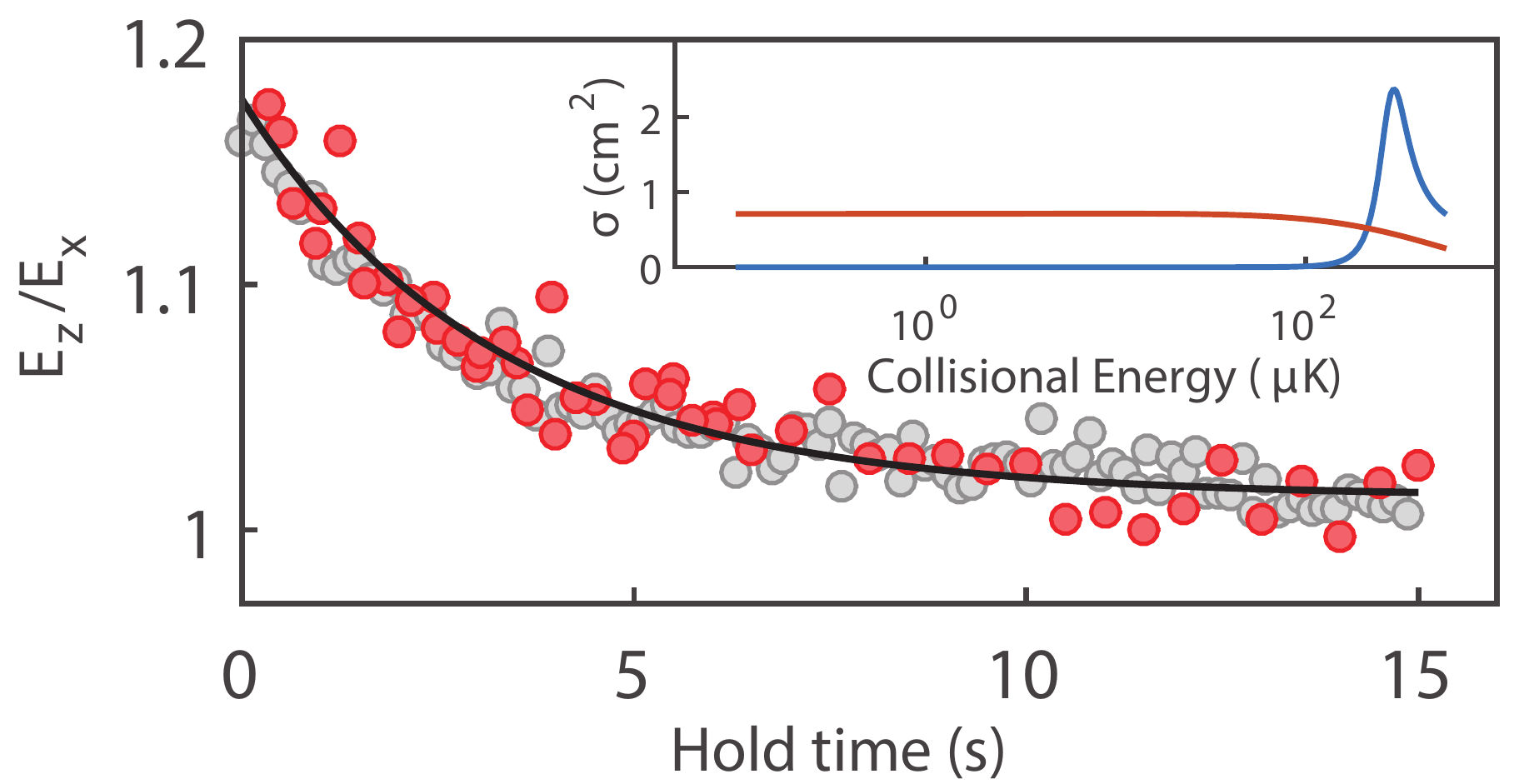}
	     \caption{Cross-dimensional relaxation of \Rb\ after introducing an anisotropy between the kinetic energies in the axial ($E_z$) and radial ($E_x$) directions.  Measured (red circles) and simulated (gray circles) ratios ${E_z}/{E_x}$ vs.\ Hold time in the $B^\prime = 328$ G/cm spherical quadrupole trap.  The inset shows the theoretically calculated \Rb- \Rb\ cross section as a function of collisional energy\cite{RbEcrossSection}. The dashed orange line represents the s wave cross section. The solid blue line represents the d wave cross section.}
\label{fig:EzExRbAspectRatio}
\end{center}
 \end{figure}

For the first, Rb-only experimental variation, we placed a small number of rubidium atoms (around $1.7 \times 10^7$) in the spherical-quadrupole trap with $B'$=328 G/cm, at a temperature around 220 $\mu$K, and observed their collisional thermalization from an initial distribution where $E_z$ is about 20$\%$ larger than $E_x$. We determined the ratio $E_z/E_x$ through time-of-flight absorption imaging after a variable time of evolution in the magnetic trap.  As shown in Fig.\ \ref{fig:EzExRbAspectRatio} for one experimental setting, the ratio is observed relax to near unity within a couple of seconds.

The measured relaxation time $\tau^{\mathrm{(ex)}}_\mathrm{Rb,Rb}$ is determined by fitting the measured $E_z/E_x$ to an equation describing exponential decay to equilibrium \cite{gold04},
\begin{equation}
\begin{aligned}
\frac{E_z}{E_x}=\frac{1+2\epsilon \exp(-t/\tau^{\mathrm{(ex)}}_\mathrm{Rb,Rb})}{1-\epsilon \exp(-t/\tau^{\mathrm{(ex)}}_\mathrm{Rb,Rb})},
\end{aligned}
\label{eq:expdecay}
\end{equation}
where $\epsilon$ quantifies the initial departure from equilibrium.  For the data of Fig.\ \ref{fig:EzExRbAspectRatio}, this relaxation time is measured to be $\tau^{\mathrm{(ex)}}_\mathrm{Rb,Rb} = 4.4(3)$ s.

In the second, Li-Rb experimental variation, we trapped both lithium and rubidium atoms in the magnetic trap.  At the much larger Rb atom numbers (about $6 \times 10^8$) used in these two-element experiments, collisions between rubidium atoms quickly evolve the rubidium gas to a state of thermal equilibrium in the trap. For \Rb, in addition, we have a microwave tone on during the experiment. However, with this microwave knife, the trap depth for \Rb\ is still deep enough so that there is no noticable forced evaporation happen, and \Rb\ stays at a constant temperature, with around 14\% decreasing in total atom number over the entire experimental window. In these experiments, we monitored the evolution of the momentum distribution of the co-trapped gas of around $2 \times 10^7$ lithium atoms.

\begin{figure}[t]
	\begin{center}
		\includegraphics[width=0.45\textwidth]{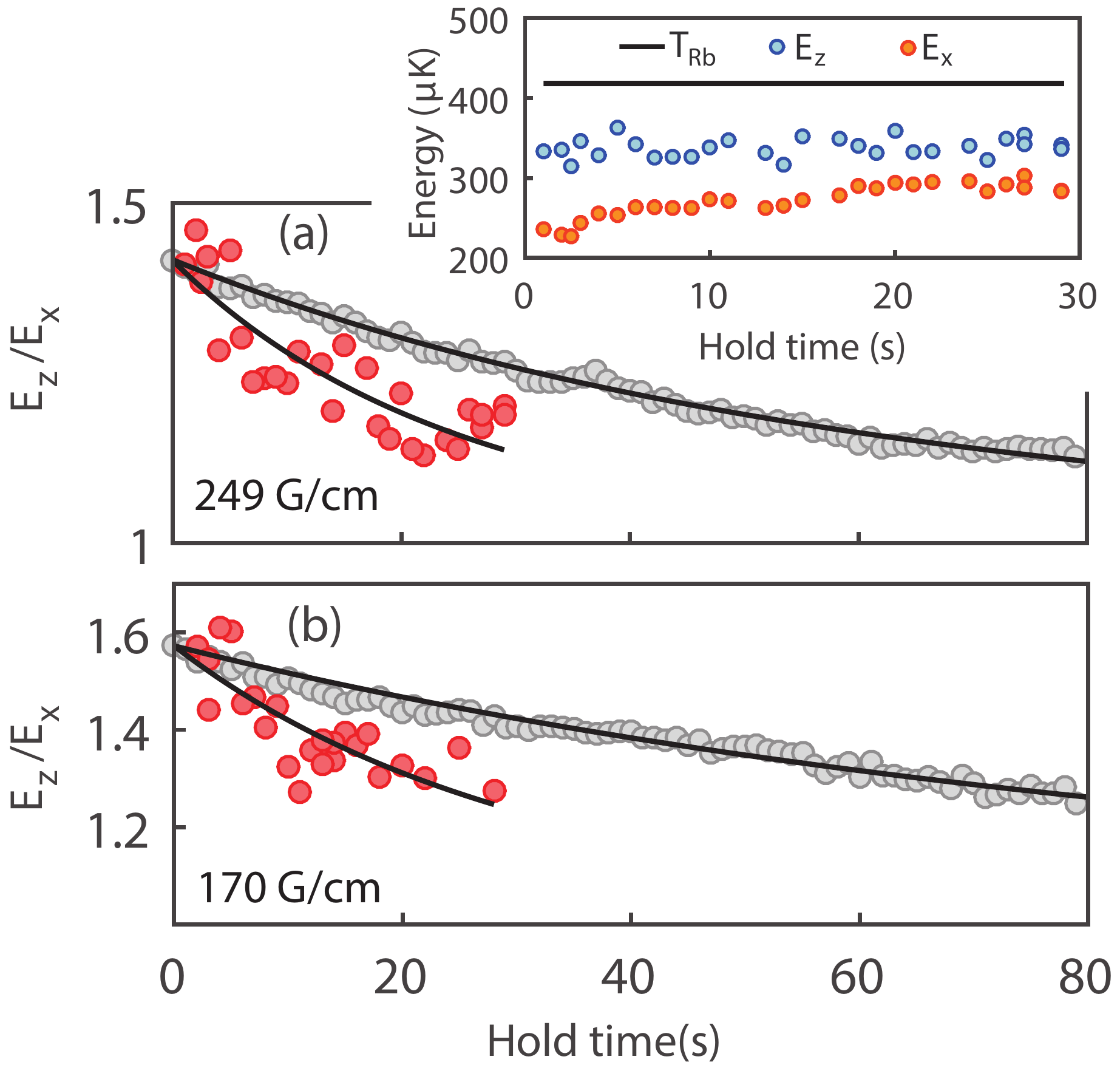}
	     \caption{Cross-dimensional relaxation of $^7$Li atoms within a co-trapped $^{87}$Rb gas.  Red circles show the measured kinetic energy ratio ($E_z/E_x$) for the lithium gas. Gray circles show results of Monte Carlo simulations accounting for experimental parameters and the theoretically predicted energy-dependent differential cross section derived using the $^7$Li-$^{87}$Rb molecular potential modeled in Ref.\ \cite{maie15efimov}.  Results at two trap strengths are shown: $B^\prime = 249$ G/cm (a) and $B^\prime = 170$ G/cm (b). Inset: neither $E_z$ (solid blue) nor $E_x$ (hollow orange), at $B^\prime = 249$ G/cm trap strength, reaches the isotropic kinetic energy of the co-trapped rubidium gas (solid black line) during the accessible evolution time.}
\label{fig:EzEx_Li}
\end{center}
 \end{figure}

In contrast with the apparent non-erodicity of the lithium-only gas, here, we observe the initially anisotropic lithium momentum distribution begin to relax toward an isotropic distribution through collisions with the rubidium atoms (Fig.\ \ref{fig:EzEx_Li}).  The cross-dimensional relaxation time of the lithium gas, $\tau^{\mathrm{(ex)}}_\mathrm{Li,Rb}$, is determined by fitting the ratio $E_z/E_x$ of the lithium gas to the same fitting function as above.  The lithium gas undergoes cross-dimensional relaxation over timescales (many seconds) that are much longer than the timescale for relaxation of the rubidium gas (100's of ms at the high Rb atom number used here).  This stark difference demonstrates that the rate of Li-Rb collisions is much smaller than that of Rb-Rb collisions.

While the lithium gas evolves toward an isotropic momentum distribution, it does not fully thermalize (i.e.\ reach a common temperature) with the rubidium gas over the timescales of our measurement.  Rather, we observe cross-dimensional relaxation to happen faster than thermalization (see Fig.\ \ref{fig:EzEx_Li} inset).  The large difference between the cross-dimensional relaxation and thermalization rates is explained by the large mass difference between lithium and rubidium atoms.  In the limit that the rubidium mass is infinite, a lithium atom undergoing an s-wave collision with a rubidium atom would emerge with a completely isotropic momentum probability distribution, so that cross-dimensional relaxation occurs in just a single collision (per atom) for a homogeneous system.  In that same collision, the lithium atom would exchange no energy with the rubidium atom, meaning that the collisions could not alter the total energy, and hence the temperature, of the lithium gas.  The true mass ratio between rubidium and lithium ($87/7$) is large, implying that cross-dimensional relaxation of lithium still occurs within the order of just one collision per atom with the rubidium gas.  In contrast, thermalization to a common temperature of the lithium and rubidium gases requires a larger number of collisions, roughly 2.7/$\xi$ = 9.8, where $\xi$ = $\frac{4m_{Li}m_{Rb}}{(m_{Li}+m_{Rb})^2}$ \cite{marz07therm}, assuming, again, s-wave interactions with constant cross section.

\section{Modeling cross-dimensional relaxation and comparison to theory}
\label{sec:modeling}

Numerical simulations are used to relate cross-dimensional relaxation to the collisions that produce such relaxation.  We first consider an approximate treatment where cross-dimensional relaxation is assumed to occur through isotropic (s-wave) collisions with an energy-independent scattering cross section $\sigma_\mathrm{cdr}$.  In this manner, we can interpret our measurement as an indirect determination of a hypothetical ``cross-dimensional relaxation cross section,'' $\sigma_\mathrm{cdr}$, which causes cross-dimensional relaxation that matches experimental measurements. The utility of this approach is that it allows us to make analytic estimates of the relation between the relaxation time $\tau$ and the cross-dimensional relaxation cross section $\sigma_\mathrm{cdr}$, and that  $\sigma_\mathrm{cdr}$  can be related intuitively to the actual theoretically predicted cross section for collisions.

As an analytic estimate, we consider that the cross-dimensional relaxation time is simply $\tau = \alpha/\Gamma$ where $\Gamma$ is the appropriate mean per-particle collision rate and $\alpha$ is a constant of order unity.  We consider small anisotropic perturbations atop the isotropic distributions of two trapped gases, with atom numbers $N_a$ and $N_b$, at the separate temperatures $T_a$ and $T_b$, respectively.  We calculate $\Gamma$ as the ensemble-averaged per atom collision rate of atoms of type $a$ with atoms of type $b$ evaluated for the isotropic momentum distributions of the trapped gases.  Integrating over velocity and position in the spherical-quadrupole trap, and neglecting gravity and Breit-Rabi corrections to the magnetic trap potential, we obtain
\begin{equation}
\begin{aligned}
\tau = \alpha
\frac{32 \pi (T_a + T_b)^3}{k^3 N_b \sigma_\mathrm{cdr} v_\mathrm{rel}}
\end{aligned}
\label{eq:analytic}
\end{equation}
where $k = \mu B^\prime / k_B$,  $v_\mathrm{rel}=\sqrt{\frac{8k_B}{\pi}\left(\frac{T_{a}}{m_{a}}+\frac{T_{b}}{m_{b}}\right)}$ is the ensemble-averaged relative velocity in a collision between atoms of type $a$ and $b$, and $m_{a,b}$ are the respective atomic masses.  We apply this expression to the rubidium-only experimental variation by assigning both the labels $a$ and $b$ to the rubidium gas, and to the lithium-rubidium experimental variation by assigning label $a$ to lithium, and label $b$ to rubidium. The constants $\alpha$ are expected to differ for these two experimental variations.  This analytic expression clarifies the functional dependence of $\tau$ on experimental parameters such as atom number, trap strength, and temperatures.


Monte Carlo numerical simulations confirm this functional dependence and provide numerical values for the constant $\alpha$ in the two experimental variations. For the rubidium-only case, $\alpha_\mathrm{Rb,Rb}$ is determined to be 2.68(6). For the lithium-rubidium case, $\alpha_\mathrm{Li,Rb}$ is determined to be 1.71(9).  Note that, in applying our analytic expression to the lithium-rubidium case, we make use of the fact, observed both experimentally and within our Monte Carlo simulations \cite{SMref}, that the cross-dimensional relaxation occurs faster than thermalization, allowing us to assign different temperatures to the lithium and rubidium gases.
The difference between $\alpha_\mathrm{Rb,Rb}$ and $\alpha_\mathrm{Li,Rb}$ is due to the different mass imbalance in the two cases.

The empirically determined  cross sections for cross-dimensional relaxation are obtained by comparing results of the Monte Carlo model to the experimentally measured cross-dimensional relaxation rates, with the results presented in Table \ref{tab:rates}. For the Rb-only measurements, we obtain $\sigma_\mathrm{cdr}^{\mathrm{(ex)}} = 7.27 \times 10^{-12} \, \mbox{cm}^2$.  This value is close to the known Rb-Rb collision cross section at the limit of zero temperature, $\sigma = 8 \pi a_\mathrm{Rb,Rb}^2 = 7.09 \times 10^{-12} \, \mbox{cm}^2$ determined from the zero-energy s-wave scattering length $a_\mathrm{Rb,Rb} = 100.40 \, a_0$ \cite{egor13} where $a_0$ is the Bohr radius.

In fact this close agreement is somewhat fortuitous.  It is known that, in $^{87}$Rb-$^{87}$Rb collisions, the s-wave scattering cross section begins to fall off for collision energies above about 100 $\mu$K, as shown in the caption of Fig.\ \ref{fig:EzExRbAspectRatio}.  However, this falling-off is compensated by the emergence of strong d-wave scattering owing to a shape resonance at a collision energy of about 275 $\mu$K \cite{boes97shape,thom04partial}.  We performed a Monte Carlo simulation that takes into account the theoretically predicted energy-dependent differential cross section, including both s-wave and d-wave collision channels, and using measured experimental settings.  Simulation predictions for the ratio $E_z/E_x$ are compared to measurements in Fig.\ \ref{fig:EzExRbAspectRatio}.  The theoretically predicted cross-dimensional relaxation time, $\tau^{\mathrm{(th)}} = 4.9(1)$ s (see Table \ref{tab:rates}), which is obtained by fitting to the Monte Carlo data agrees with the experimental measurement within about 10\%. The good agreement between our measurements and simulation demonstrates the validity of our method for measuring the collision cross section through cross-dimensional relaxation.


The Li-Rb measurements were performed at two different trap strengths. When accounting for the different experimental settings at which measurements are performed, we obtain two values for $\sigma^{\mathrm{(ex)}}_\mathrm{cdr}$ that agree with one another within the estimated error, using Eq.\ \ref{eq:analytic} to propagate statistical and systematic errors in our measurements of temperature, trap strength, and atom number (see Supplemental Material \cite{SMref}).


\begin{table*}[t]
\resizebox{\textwidth}{!}{
\begin{tabular}{|c|c|c|c|c|| c|c|c || c|c |}
  \hline
  $B^\prime$ & $N_{Li}$ & $T_{Li}$ & $N_{Rb}$ & $T_{Rb}$ & $\tau^{\mathrm{(ex)}}$ & $\sigma^{\mathrm{(ex)}}_\mathrm{cdr}$ &$\sigma^{\mathrm{(ex)}}_\mathrm{T}$ &$\tau^\mathrm{(th)}$ & $\sigma^\mathrm{(th)}_\mathrm{cdr}$ \\
  \hhline{|=|=|=|=|=#==|=#=|=|}
  328 G/cm & -- & -- & $1.66(6) \times 10^7$ & 223(2) $\mu$K & 4.4(3) s & $7.3(6) \times 10^{-12} \, \mbox{cm}^2$& -- & 4.9(1) s & $6.6(3) \times 10^{-12} \, \mbox{cm}^2$ \\
  \hhline{|=|=|=|=|=#==|=#=|=|}
  249 G/cm & $2 \times 10^7$ & 290(9) $\mu$K & $6.4(1) \times 10^8$ & 418(4) $\mu$K & 28(4) s & $8.5(1.1) \times 10^{-14} \, \mbox{cm}^2$&$8.6(1.2) \times 10^{-14} \, \mbox{cm}^2$ & 69(1) s & $3.5(1) \times 10^{-14} \, \mbox{cm}^2$ \\
  170 G/cm & $2 \times 10^7$ & 243(6) $\mu$K & $5.8(2) \times 10^8$ & 300(2) $\mu$K & 37(6) s & $10.3(1.8) \times 10^{-14} \, \mbox{cm}^2$ & $12.3(2.0) \times 10^{-14} \, \mbox{cm}^2$ & 95(2) s & $4.0(2) \times 10^{-14} \, \mbox{cm}^2$ \\
  \hline
\end{tabular}}
\caption{Measured cross-dimensional relaxation rates $\tau^{\mathrm{(ex)}}$ and comparisons to theory.  Experimental settings are characterized by different trapping magnetic field gradients $B^\prime$, initial trapped atom numbers $N_{Li}$ and $N_{Rb}$, and measured temperatures upon cross-dimensional relaxation $T_{Li}$ and $T_{Rb}$.  For Rb-only experiments, we include only information on the Rb gas.  Monte Carlo simulations for atoms interacting with the theoretically predicted differential cross section, under conditions matching the experimental settings, are used to determine a theoretically predicted cross-dimensional relaxation time $\tau^\mathrm{(th)}$. Separate Monte Carlo simulations for atoms interacting via energy-independent s-wave cross section $\sigma_{\mathrm{cdr}}$, under conditions matching the experimental setting, are used to determine a linear relationship between $\sigma_\mathrm{cdr}^{-1}$ and relaxation time(Fig.\ \ref{fig:crossSection_QDT} inset). This relation is used to drive $\sigma_\mathrm{cdr}^{(ex)}$ from $\tau^\mathrm{(ex)}$ and $\sigma_\mathrm{cdr}^\mathrm{(th)}$ from $\tau^\mathrm{(th)}$. $\sigma_\mathrm{T}^{(ex)}$ is the implied experimental value from the thermally averaged cross section, see Section \ \ref{sec:theory} for details. All one-sigma error estimates are for statistical errors.  Systematic errors in determining experimental settings yield systematic errors at the level of around 5\% for $\sigma^{\mathrm{(ex)}}_\mathrm{cdr}$, where the systematic error related to \Rb\ atom number calibration is not included. \Rb\ atom number calibration is constrained by the \Rb\ -only experiment to 10\% percent level. \label{tab:rates}}
\end{table*}

\begin{figure}[t]
	\begin{center}
		\includegraphics[width=0.45\textwidth]{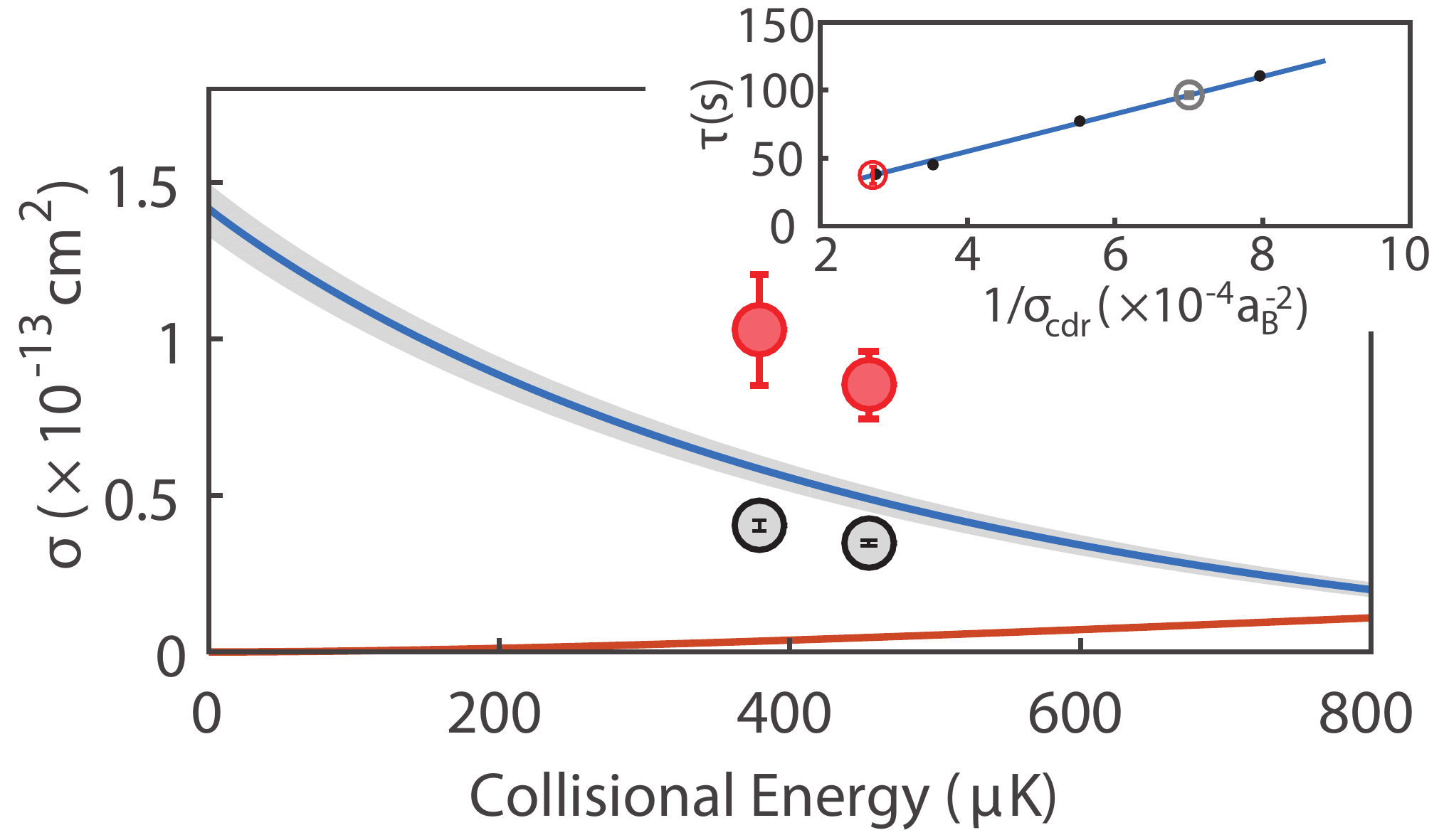}
	     \caption{Comparison between experimental measurements and theoretically predicted cross section for $^7$Li-$^{87}$Rb collisions. The solid blue (dashed orange) lines give the s-wave (p-wave) cross section calculated using the Maier \emph{et al.}\ model potential. The shaded gray area represents the smallest and largest theoretical allowable s-wave cross section with $\chi^2_5 < 20$ (see Sec.\ \ref{sec:theory}).  For comparison, we show $\sigma_\mathrm{cdr}$ obtained either experimentally (red circles(upper), with one-sigma statistical error bars) or theoretically (gray circles(lower), based on the Maier \emph{et al.}\ model potential).  These are plotted at the thermally averaged collision energies determined by experimental conditions.  The corresponding cross-dimensional relaxation times derived from experiment and theory are shown in the insert (same color scheme).  We show also Monte Carlo simulated results (black symbols) under experimental conditions using a series of energy-independent isotropic cross sections, from which a linear relationship between $\sigma_\mathrm{cdr}^{-1}$ and the cross-dimensional relaxation time $\tau$ is obtained.}
\label{fig:crossSection_QDT}
\end{center}
 \end{figure}

We compare these measurements to Monte Carlo numerical simulations that account for both the anisotropy and the energy dependence of the differential collision cross section. This differential cross section includes contributions from both s and p partial waves; higher partial waves are irrelevant at the collision energies of our experiments \cite{pere15qdt}.  Our simulations take into account the slight Rb atom loss observed in our experiments over the long measurement times.  The result of this comparison is shown both in Table \ref{tab:rates} and Fig.\ \ref{fig:crossSection_QDT}.

This comparison highlights the significant deviation between our experimental findings and theoretical predictions.  In brief, we observe Li-Rb collisions to cause cross-dimensional relaxation at a rate that is 2.5 times faster than is predicted by simulations based on the predicted differential cross section.  Equivalently, we observe a cross section for cross-dimensional relaxation that is 2.5 times larger than predicted theoretically.

We note that the energy anisotropy $E_z/E_x$ should not approach unity through purely exponential decay (i.e.\ following Eq.\ \ref{eq:expdecay}). There exist particle trajectories for trapped lithium atoms, e.g.\ those with large axial angular momentum, whose trajectory-averaged rubidium gas density is lower than the volume average.  Lithium atoms in such trajectories collide less frequently than those that experience higher rubidium gas density.  Non-exponential decay may be further enhanced if the cross-section has a rapid energy dependence, such as that predicted by theory (Fig.\ \ref{fig:crossSection_QDT}).  We find evidence of the resulting non-exponential decay of the momentum distribution toward isotropy in numerical simulations, where we find that the fitted value for $\tau$ varies with the range of simulated data used for the fit, with later-time data giving systematically longer relaxation times. However, our experimental data are too noisy to confirm this non-exponential decay. To make direct comparison between experiment and theory, we compare $\tau^{ex}$ to $\tau^{th}$ with a fitting window $t_f = 28 s$, corresponding to the entire experimental time window. Nevertheless, the experimentally observed relaxation occurs always significantly faster than theoretically predicted.

\section{Theoretical Investigations of the cross section}
\label{sec:theory}

This section describes the explorations we have carried out to modify the theoretical description of two-body Li-Rb scattering in a way that preserves existing agreement between theory and experimentally measured Fano-Feshbach resonance properties, while also agreeing with the present experimental results.  The modified theoretical model considered in this study starts with the LiRb singlet and triplet potential energy functions that were previously optimized by Maier \emph{et al.}\ \cite{maie15efimov}.  Our approach is to modify this potential, as described below, and then to solve the coupled differential equations relevant to cold collisions with a finite element method (FEM)-R matrix calculation \cite{burk97}.  We explored but do not rely here upon approximate quantum defect theory treatments \cite{burk98,ruzi13,pere15qdt}.

We confirm that our calculation method is valid by using the Maier \emph{et al.}\ model potentials and reproducing the calculated positions of several reported s-, p- and d-wave Fano-Feshbach resonances in $^7$Li-$^{87}$Rb.  
The agreement between these theoretical predictions and experimental findings is quantified by calculating $\chi^2_5 = \sum \frac{(B_{\mathrm{calc}} - B_{\mathrm{exp}})^2}{(\delta B)^2}$ where the sum is taken over the five measured Fano-Feshbach resonances, with $B_{\mathrm{calc}}$ being the theoretically predicted and $B_{\mathrm{exp}}$ the experimentally measured magnetic field position of the resonance, and $\delta B = 0.5$ G being the experimentally reported uncertainty.  For the Maier \emph{et al.}\ model potential, which was selected to match all experimental data on the LiRb potential at the time of its publication (obviously excluding the present measurements), one finds $\chi^2_5 = 11$.  The Maier \emph{et al.}\ model potential is used to derive the s- and p-wave collision cross sections, as functions of the collision energy, which are presented in Fig.\ \ref{fig:crossSection_QDT}.  As discussed previously, this energy-dependent differential cross section is used within Monte Carlo simulations to derive a cross-dimensional relaxation cross section (also in Fig.\ \ref{fig:crossSection_QDT}) that is significantly smaller than that determined by measurement.  

We explored the elasticity of the model predictions to variations in several model parameters, i.e.\ we explore to what extent the collision cross section between $^7$Li and $^{87}$Rb, with both atoms in the $|F=1, m_F = -1\rangle$ hyperfine state, can vary while maintaining agreement also with the measured Fano-Feshbach resonance positions.  For this, we consider three modifications of the two-body potential: a modification of the inner part of the spin-singlet potential ($X ^1\Sigma^+$), a modification of the inner part of the spin-triplet potential ($a ^3\Sigma^+$), and a modification of the $C_6$ coefficient that characterizes the long-range van der Waals interaction.  After making such modifications, we recompute the predicted positions of the five Fano-Feshbach resonances, for comparison with the Maier \emph{et al.}\ measurements \cite{maie15efimov}.

We also compute the energy dependent differential cross section with both atoms in the $|F=1, m_F = -1\rangle$ and at low magnetic field\cite{LiRbBCrossSection}. In order to compare this energy-dependent function to our experimental findings, rather than resorting in all cases to detailed Monte Carlo calculations of cross-dimensional relaxation, we compute instead the thermally averaged cross section $\sigma_T$.  For this, we assume Li and Rb gases that are at uniform density and that are each at thermal equilibrium at the temperatures $T_{Li}$ and $T_{Rb}$ that are indicated in Table \ref{tab:rates}.  For example, the thermally averaged cross sections found in this manner, using the Maier \emph{et al.}\ potential, are shown also in Table \ref{tab:rates}. For this estimation, we only include s wave contribution. We compare these thermally averaged cross sections to an implied experimental value, $\sigma_T^{\mathrm(ex)}$, which we obtain by scaling up the result from the Maier \emph{et al.}\ model by the ratio $\tau^\mathrm{th}/\tau^\mathrm{ex}$ in that Table.

We consider modifications to the Li-Rb molecular potentials that remain somewhat consistent with the measured positions of the Fano-Feshbach resonances, where we qualify a modification as consistent so long as $\chi^2_5$ remains of the same order as its value for the Maier \emph{et al.}\ potential.  Specifically, by assessing how the inferred Fano-Feshbach resonance positions and also the thermally averaged cross-section vary linearly with the three potential model modifications, we determine model settings that produce the largest and smallest $\sigma_T$ within the space bounded by $\chi^2_5 < 20$(Fig.\ \ref{fig:crossSection_QDT} shaded area). We compute the full energy-dependent cross-section for these two model settings.  As shown in Fig.\ \ref{fig:crossSection_QDT}, constraining the model to remain consistent with the measured positions of the Fano-Feshbach resonances permits only slight modifications, at the level of less than 10\%, to the predicted collision cross section and, similarly, to the predicted cross-dimensional relaxation rate.  These modifications are insufficient to bridge the difference between our experimental findings and theoretical predictions.  Overall, we conclude that the measured Fano-Feshbach resonance potentials and our measured cross-dimensional relaxation rate cannot all be made simultaneously consistent with present-day models of the Li-Rb molecular potential.

\section{Conclusion}



We have characterized \Li-\Rb\ collisions at a collision energy that is in the range of several 100's of $\mu$K by observing cross-dimensional relaxation of gases trapped in a spherical quadrupole magnetic trap.  To our knowledge, ours is the first use of a spherical quadrupole trap in a two-species cross-dimensional relaxation measurement.  Our experimental method makes use of several specific properties of the two elements that we have chosen to study.  In this specific instance, one of the elements (element $a$; here $^7$Li) has very weak interactions and is very dilute, so that the collision rate $\Gamma_{aa}$ within the single-element gas is small.  For this reason, loaded on its own into a spherical quadrupole trap, this gas evolves non-ergodically for very long time, as we have confirmed experimentally.  The other element (element $b$; here $^{87}$Rb) has higher density and stronger interactions, so that the collision rate $\Gamma_{bb}$ in a single-element gas is large.  Thus, this gas thermalizes rapidly on its own in the magnetic trap.  The interaction between the two elements, characterized by collision rate $\Gamma_{ab}$, is intermediate between $\Gamma_{aa}$ and $\Gamma_{bb}$.  This fact allows us to ascribe the cross-dimensional relaxation of element $a$ in the presence of element $b$ as being solely due to $a$-$b$ collisions, and, also, to treat the relaxation process quantitatively by assuming element $b$ is always at thermal equilibrium. Another specific feature of our experiment is the large mass ratio $m_{Rb}/m_{Li}$, owing to which the lithium gas undergoes cross-dimensional relaxation well before it undergoes full thermalization with the rubidium gas with which it interacts.  Our measurements show clearly the difference between the cross-dimensional relaxation and thermalization timescales.

We measure the cross-dimensional relaxation rate for the $^7$Li gas in the presence of $^{87}$Rb under two experimental settings.  These measured rates disagree significantly with theoretical predictions based on Monte Carlo simulations of cross-dimensional relaxation under our experimental conditions, and on the theoretically predicted, energy-dependent different cross section for Li-Rb collisions.  A careful theoretical investigation shows that it is hard to incorporate our results to the existing \Li\-\Rb\ model potentials without inducing a disagreement between predicted and measured positions of several Fano-Feshbach resonances.

Future investigations to address this disagreement are warranted.  It may be valuable to characterize elastic collisions among atoms in the $|F=1, m_f = -1\rangle$ atoms of $^7$Li and $^{87}$Rb over a wider range of collision energies.  This can be done by repeating cross-dimensional relaxation experiments at various gas temperatures, or by characterizing cold collisions at well-defined and variable collision energy \cite{thom04partial,bugg04dwave,gens12}.  Such measurements could confirm the sharp energy dependence of the cross section, which is predicted theoretically, with the exact energy of the zero-crossing of the s-wave scattering phase providing a very tight constraint for theoretical models.  It will also be valuable to examine spin-dependent interactions between the $F=1$ $^7$Li and $^{87}$Rb spinor-gas systems, providing more information on both the singlet and triplet scattering lengths for heteronuclear collisions.  Experimental measurements on Fano-Feshbach resonances in $^7$Li-$^{87}$Rb collisions could be refined by applying methods such as rf-modulation or interferometric measurements \cite{chin10rmp} to identify and measure the energies of shallow molecular bound states.  Such experiments could supplement the existing measurements \cite{maie15efimov} to determine the resonance positions with more certainty.

We thank E.\ Tiemann for providing the Born-Oppenheimer potentials that were previously optimized by his group, and C.\ Zimmermann for further information on the Fano-Feshbach resonance measurements reported in Ref.\ \cite{maie15efimov}.  The work of CHG and YW has been supported by National Science Foundation grant No.\ PHY-1912350.  The work of FF, SW, JI, AS, and DSK has been supported by National Science Foundation grant No.\ PHY-1707756, by DTRA and by NASA.

\bibliography{lirb_notes,allrefs_x2}

\begin{thebibliography}{41}
\bibitem [{CDR()}]{CDRClarify}%
  \BibitemOpen
  \href@noop {} {}\bibinfo {note} {The words "cross-dimensional relaxation"
  refer to "x - z" kinetic energy relaxation. In other cases, people might
  refer to experiments where two species are trapped in different spatial
  dimension, which is not the case we are studying here.}\BibitemShut {Stop}%
\bibitem [{Clo()}]{CloudSizeClarify}%
  \BibitemOpen
  \href@noop {} {}\bibinfo {note} {The spatial extent can be estimated as
  $l_{cloud} = \frac{k_BT}{\mu B'}$. Here $k_B$ is boltzmann constant, $T$ is
  the gas temperature, $B'$ is the magnetic field gradient along the trap axis,
  and $\mu = 0.5\mu_B$ is atomic magnetic moment at low field with $\mu_B$ the
  Bohr magneton. A quick estimation for the Li gas trapped in a $B^\prime$ =
  240 G/cm trap and a temperature of 290 $\mu$K shows $l_z \approx 0.7$ mm
  along the trap axis. For the Rb gas with $B'$ = 240 G/cm and $T = 418 \,
  \mu$K, this estimation shows $l_z \approx 1$ mm. We note that for the Li gas,
  a fuller account of the magnetic potential, using the Breit-Rabi expression,
  is required for a more precise determination of the spatial distribution both
  at equilibrium and in the non-equilibrium steady-state. This fuller
  treatment, which indicates the Li gas to be larger than estimated just based
  on the linear Zeeman energy, leads to a reduction in the average Rb density
  seen by Li atoms, and also affects our time-of-flight measurements of the
  kinetic energy.}\BibitemShut {Stop}%
\bibitem [{LiR()}]{LiRbBCrossSection}%
  \BibitemOpen
  \href@noop {} {\bibinfo  {journal} {Our theoretical investigation shows the
  cross section only changes about 10\% up to 60 G, which covers the entire
  sampling range of our experiment}\ }\BibitemShut {NoStop}%
\end{thebibliography}


\end{document}